# Particle-hole Hall effect in graphenelike structures

J. C. Martinez[1], M. B. A. Jalil

*Information Storage Materials Laboratory, Electrical and Computer Engineering Department, National University of Singapore, 4 Engineering Drive 3, Singapore 117576*

S. G. Tan

*Data Storage Institute, DSI Building, 5 Engineering Drive 1, National University of Singapore, Singapore 117608*

We show that a moderately strong constant electric field in the plane of a monolayer graphene sheet can create particle-hole pairs at an observable rate. The pairs undergo zitterbewegung in opposite directions leading to a Hall-like separation of the charge carriers and a measurable transverse dipole moment is predicted which serves as the signature of the zitterbewegung. In contrast with the created pairs, the zero modes of the excitation induce a current transverse to the electric field but do not result in separated charges. For bilayer graphene a similar effect by the electric field is shown not to be possible.

[1] Corresponding author email: elemjc@nus.edu.sg

For heavy nuclei the binding energy of an electron in the lowest orbital grows faster than the square of the atomic number $Z$. If $Z = 145$ the binding energy of this electron exceeds the mass of an electron so adding an electron diminishes the atom's mass. When $Z = 173$ the binding energy equals the mass of an electron-position pair so the creation of such a pair requires no expenditure of energy [1]. If $Z$ is even greater so the Coulomb field is the more intense, pair production could occur spontaneously and would then be detected by the emergence of the positron expelled from the nucleus and the decrease of the nuclear charge because of the residual electron bound to the innermost orbital. In turn this implies a 'charged' vacuum near the nucleus whose properties differ markedly from those of an ordinary uncharged vacuum [1, 2]. This effect whereby a super-heavy atomic nucleus becomes spontaneously unstable has not been seen because the formation of such nucleus is far beyond present-day experimental realization. One wonders whether the ultimate flat sheet graphene, due to its unusual Dirac-Weyl type band-structure near the Fermi level where the Fermi surface reduces to two inequivalent K points of the Brillouin zone, might not be a candidate in which an analogous effect could occur [3]. Although we do not give an unqualified affirmative answer we note nevertheless that a new particle-hole Hall effect exists for Dirac particles in monolayer graphene which may potentially be observed with available technology.

We will show that a moderately strong uniform static electric field (analogous in this case to the Coulomb field of a large $Z$ nucleus) in the plane of a graphene sheet can induce the creation of a particle-hole pair (or pair for short) per unit cell of the graphene lattice [4]. As each pair is pulled apart by the electric field, the particle and hole undergo zitterbewegung in opposite directions evolving ultimately into separated charge carriers and, on a macroscopic level, to the formation of a polarized sheet which serves as the signature of the pairs produced. Such sheets might find application in the formation of fullerenes and carbon nanotubes [5]. We estimate the dipole moment that can be expected which turns out to be of the same order as that of a boron nitride nanotube [6]. The measurable dipolar polarization can thus serve as a signature of the pair-production process in graphene following from a Hall-like charge separation. Moreover the experimental investigation of zitterbewegung is the object of current interest because it is thought that in quantum wires it could occur in the terahertz regime [7, 8]; its dynamical nature which we consider is also of importance in femtosecond pulse technology [9]. We note that an analogous effect cannot be induced by an electric field in bilayer graphene. Additionally, graphene is a gapless semiconductor whose band Hamiltonian is identical to the zero-mass limit of the Dirac equation, so it will come as no surprise that the massless pair produced illustrates the axial anomaly in which parity is violated [10, 11]. In verifying this we find that an induced current appears as a result of the polarization of the Dirac sea in response to the electric field but which does *not* lead to separated charges. This phenomenon is attributed to the zero modes of the system whilst the aforementioned pair production and separation are traced to the non-zero modes.

Consider the low-energy excitations with momenta in the vicinity of the Fermi level near the K point in the Brillouin zone of graphene which obey a Dirac-type equation [10]

$$\begin{pmatrix} i\partial_t + qE_0 y - \Delta & \partial_x - i\partial_y \\ \partial_x + i\partial_y & -(i\partial_t + qE_0 y + \Delta) \end{pmatrix} \begin{pmatrix} \phi(x,y,t) \\ \chi(x,y,t) \end{pmatrix} = 0 \qquad (1)$$

where $v_F$ (Fermi velocity) and $\hbar$ are set to unity and a static uniform electric field $-E_0$ in the *y*-direction is present. We are working in the space-dependent Coulomb gauge so the vector potential is $A_\mu = (E_0 y, 0, 0)$. The matrices operate on the one-valley, triangular sublattice (pseudospin) space of the graphene honeycomb structure corresponding to the A and B atoms. Although the fermions are massless we have introduced a $\Delta$ term for later convenience. Each Fourier mode of the field can be expanded as $\phi(x,y,t) = e^{i(k_x x - \omega t)} a(y)$, $\chi(x,y,t) = e^{i(k_x x - \omega t)} b(y)$ so Eq. (1) simplifies into two Schrodinger-type equations for a unit mass in an inverted harmonic oscillator potential

$$\left( \frac{d^2}{d\xi^2} + \frac{1}{4}\xi^2 - a_{k\perp} \pm \frac{i}{2} \right) \binom{a(\xi)}{b(\xi)} = 0 \qquad (2)$$

in which $\xi = \sqrt{2/qE_0}\,(\omega + qE_0 y)$, $a_{k\perp} = (k_x^2 + \Delta^2)/2qE_0$. Since the particle energy corresponding to Eq. (2) is negative, the above describes tunneling from a turning point $y_-$ to another turning point $y_+$. The positive-energy solutions can be given in terms of complex parabolic cylinder functions [12]

$$a \pm b \propto E(a_{k\perp} \mp \tfrac{i}{2}, \xi) \to A \frac{1}{|\xi|^{\frac{1}{2} \pm \frac{1}{2}}} \varphi_{k_x}(\xi) \pm B \frac{1}{|\xi|^{\frac{1}{2} \pm \frac{1}{2}}} \varphi^*_{k_x}(\xi), \quad \xi \ll -2\sqrt{|a_{k\perp} \pm \tfrac{i}{2}|}$$
$$\to C \frac{1}{|\xi|^{\frac{1}{2} \pm \frac{1}{2}}} \varphi^*_{k_x}(\xi), \quad \xi \gg 2\sqrt{|a_{k\perp} \pm \tfrac{i}{2}|} \qquad (3)$$

where $A = i\sqrt{2(e^{2\pi a_{k\perp}} - 1)}$, $B = \sqrt{2} e^{\pi a_{k\perp}}$, $C = \sqrt{2}$, $\varphi_{k_x}(\xi) = e^{-i\xi^2/4} e^{i\alpha}$, $\alpha = a_{k\perp} \ln \xi - \tfrac{1}{2}\phi_2 - \tfrac{\pi}{4}$ and $\phi_2 = \arg \Gamma(\tfrac{1}{2} \pm \tfrac{1}{2} + ia_{k\perp})$, and we have also given the asymptotic forms far away from $\xi = 0$. We observe that $\varphi^*_{k_x} e^{-i\omega t}$ describes an incoming particle from $\xi = -\infty$ and an outgoing particle to $\xi = +\infty$. Similarly $\varphi_{k_x} e^{-i\omega t}$ describes outgoing particles to $\xi = \pm\infty$.

By reversing the signs of $\omega$ and $q$ in Eq. (1) we can obtain similar results for the negative-energy solutions:

$$\bar{a} \pm \bar{b} \propto E(a_{k_\perp} \pm \tfrac{i}{2}, -\xi) \to -iA \frac{1}{|\xi|^{\frac{1}{2} \mp \frac{1}{2}}} \varphi_{k_x}(\xi) \mp B \frac{1}{|\xi|^{\frac{1}{2} \mp \frac{1}{2}}} \varphi^*_{k_x}(\xi), \quad \xi \gg 2\sqrt{\left|a_{k_\perp} \pm \tfrac{i}{2}\right|}$$
$$\to C \frac{1}{|\xi|^{\frac{1}{2} \mp \frac{1}{2}}} \varphi^*_{k_x}(\xi), \quad \xi \ll -2\sqrt{\left|a_{k_\perp} \pm \tfrac{i}{2}\right|} \tag{4}$$

in which the over-bars distinguish this case from the positive-energy solutions and the symbols have the same meanings here. Now, however, we have to interchange the interpretations of $\varphi^*_{k_x} e^{-i\omega t}$ and $\varphi_{k_x} e^{-i\omega t}$, and understand $e^{-i\omega t}$ to mean $e^{i|\omega|t}$. Thus every outgoing particle solution is matched by an incoming negative energy solution which propagates with opposite charge in the reverse direction. This symmetry is lost, however, for the zero modes and this spectral asymmetry will be discussed later. We interpret $|C/B|^2$ as the tunneling probability and $1 - |C/B|^2 = |A/B|^2$ as the reflection probability or vacuum-vacuum probability, consistent with flux conservation [13, 14].

The interpretation above leads to an expression for the pair-production rate if we appeal to an analogy with instantons through potential barriers. The tunneling probability per mode $|C/B|^2 = e^{-2S_{k_x}}$ is determined by the single-instanton action

$$S_{k_x} = \int_{y_-}^{y_+} \sqrt{k_x^2 + \Delta^2 - (\omega + qE_0 y)^2}\, dy = \pi a_{k_\perp} \tag{5}$$

where $y_\pm$ are the classical turning points [15]. The single instanton and multi-instantons (corresponding to multi-traversals between turning points) have been found to be related with one-pair and multipair production, with analogous relations between anti-multi-instantons and the annihilation of created pairs. When there is no tunneling instanton, neither is there pair production. For the case at hand which involves fermions, multiparticle production is however prevented by Pauli blocking. Hence the tunneling probability given above is also the total tunneling probability. Then the relative probability for no-pair production is $1 - e^{-2S_{k_x}}$ and the fermion pair production per unit area per unit time is

$$\Gamma = 2\,\text{Im}\,\mathcal{L}_{eff}^{f} = -\frac{1}{VT}\sum_{\text{all states}}\ln(1 - e^{-2S_{k_x}}) = \frac{1}{VT}\sum_{\text{all states}}\sum_{n}\frac{1}{n}(e^{-2S_{k_x}})^n \tag{6}$$

We evaluate this to obtain the pair production rate per unit area for spin-1/2 fermions (we reinstate $\hbar$ and $v_F$ in this result) [13]

$$\Gamma = \frac{v_F}{4\pi^2}\left(\frac{qE_0}{\hbar v_F}\right)^{3/2}\sum_{n=1}^{\infty}\frac{1}{n^{3/2}}e^{-\pi\Delta^2 v_F^3 n/qE_0} \tag{7}$$

For a typical electric field of $10^7\,\text{V m}^{-1}$ and taking $v_F \approx 10^6\,\text{m s}^{-1}$ we obtain a pair production rate $\Gamma$ in graphene in the limit $\Delta = 0$ of about 1 pair per 0.25 ns per cell (area $\approx 5\times 10^{-20}\,\text{m}^2$). The effect of a spin-orbit interaction, if it exists, is to include a constant $\Delta$ in Eq. (1). However $\Delta$ is of order meV and its effect on the production rate $\Gamma$ is hardly noticeable. After the creation of a pair, it is unlikely a second one will emerge because the created pair would set up a reactive field sufficient to oppose the applied external electric field. Moreover the striking band structure in graphene holds only at half-filling when each site of the lattice yields one electron to the Fermi sea [16].

A brief digression, invoking the above developments, will show that pair production cannot occur for bilayer graphene. Still working in the Coulomb gauge Eq. (1) is now replaced by ($m$ is the corresponding fermion mass) [17]

$$(i\partial_t + qE_0 y)\Psi = \frac{\hbar^2}{2m}\begin{pmatrix} 0 & (\partial_x - i\partial_y)^2 \\ (\partial_x + i\partial_y)^2 & 0 \end{pmatrix}\Psi \tag{8}$$

We follow the same procedure as before and for simplicity choose $k_x = 0$ so there is no motion in the $x$-direction. Then the mode expansion employed in Eq. (1) yields a single Schrodinger equation $\{-\frac{\hbar^2}{2m}\frac{\partial^2}{\partial y^2} + qE_0 y\}\phi = -\hbar\omega\phi$, with $\phi = -\chi$. This is the equation for tunneling through a linear potential $qE_0 y$ from some finite turning point $y_0$ to $\infty$. The action $S_{k_x}$ in this case is infinite, that is, there are no finite instantons available. Thus all would-be instantons from one spatial infinity to another are infinite and do not contribute to the tunneling probability; hence there is no pair production in bilayer graphene.

We return to the fate of the created pair. The pair is subject to the electric field and each component undergoes Zitterbewegung or trembling motion as depicted in Fig. 1 [8, 18]. To investigate this it is convenient to choose the gauge $A_\mu = (0,0, A(t))$, where $A(t) = -\int E(t)dt$. The corresponding two-dimensional Hamiltonian in this new gauge is $\hat{H} = v_F \vec{\sigma} \cdot \vec{\Pi}$, in which $\vec{\sigma}$ are the usual Pauli matrices $(\sigma_x, \sigma_y, \sigma_z)$ and $\vec{\Pi}$ are the conjugate momenta $(p_x, p_y + qE_0 t, 0)$ (the notation refers to Cartesian vectors). We evaluate the matrix-valued velocity operator in the Heisenberg picture, $\hat{v}(t) =$ $\exp(i\hat{H}t/\hbar)\hat{v}\exp(-i\hat{H}t/\hbar)$ where $\hat{v} = v_F(\sigma_x, \sigma_y)$ is the time-independent velocity operator in the Schrodinger picture. The result can be expressed succinctly as

$$\hat{v}(t) = v_F\left(\vec{\sigma} - \frac{\vec{\sigma}\times\vec{\Pi}}{\Pi}\sin(2v_F kt) - \frac{(\vec{\sigma}\times\vec{\Pi})\times\vec{\Pi}}{\Pi^2}\{1-\cos(2v_F kt)\}\right) \quad (9)$$

where $\Pi \equiv \sqrt{\vec{\Pi}\cdot\vec{\Pi}}$ and $k \equiv \Pi/\hbar$ (in evaluating the vector products we retain in the end only the planar components). The above describes undamped trembling motion with the (time-dependent) frequency $2v_F k$, which is determined by the difference in energy between the particle and hole components (for a given $k$). For a fixed direction of $E_0$ in the plane, this trembling motion is transverse to it. If initially there is negligible motion so $\Pi$ is small to start with, then only $\Pi_y$ counts and the zitterbewegung will be in the $x$-direction superimposed over the motion along the electric field.

We evaluate the zitterbewegung assuming a created charge carrier given by a two-dimensional Gaussian wave-packet initially centered at $k_0$ with characteristic width $d$,

$$\psi(r,0) = \frac{d}{2\pi^{3/2}}\int d^2k e^{-\frac{1}{2}d^2(k-k_0)^2} e^{ik\cdot r}\begin{pmatrix}1\\0\end{pmatrix} \quad (10)$$

In undertaking this calculation we note that (a) we are working in pseudospin space, and (b) the calculation of fluctuations of an infinite plane wave is hardly meaningful as pointed out by Lock [19] and by Rusin and Zawadzki [8]. With $\psi(r,0)$ we calculate the 11 component of $\hat{v}(t)$, assuming $k_{0x} = 0$ for simplicity,

$$\langle\psi|\hat{v}_x|\psi\rangle = 2v_F e^{-d^2(k_{0y}+qE_0t/\hbar)^2} \sum_{m=1}^{\infty} \frac{[d(k_{0y}+qE_0t/\hbar)]^{2m-1}}{m!(m-1)!} \int_0^{\infty} Q^{2m} e^{-Q^2} \sin\left(\frac{2v_F t}{d}Q\right) dQ$$

(11)

We also find that $\langle\psi|\hat{v}_y|\psi\rangle = 0$. When $k_0 = 0$ and $E = 0$ then $\langle\psi|\hat{v}|\psi\rangle = 0$. Unlike Eq. (9) the result (11) shows a damped transient behavior (independent of $v_F$), consistent with zitterbewegung being a dynamical phenomenon and not a stationary one. One can also verify directly that zitterbewegung is the result of interference between the particle and hole eigenstates [8]. Results for $\langle v_x \rangle$ versus time are plotted in Fig. 2 for several realistic widths: $E_0 = 10^7 V\ m^{-1}$, $k_{0y} = 10^9 m^{-1}$, $d = 5$ (orange), 10 (dashed, blue), 20 (red) Å. The particle displacements $x_\infty$ are 0.3 and 0.45nm, respectively, for the latter two values of $d$ (see Fig. 1). This corresponds to a dipole separation of nearly 1 nm per unit cell which is not very different from that observed for BN nanotubes [6]. Thus, this predicted dipole moment can provide an experimental basis for detecting the pair production process and subsequent Hall charge separation in monolayer graphene.

We return to the zero (zero-energy) modes and show that they induce a vacuum Hall effect, but do *not* yield the separated charge carriers, unlike the non-zero energy modes as explained above. First we note the spectral asymmetry in the solutions (3) and (4): although every particle is matched by a hole, this symmetry is lost for the zero modes. To show this we multiply Eq. (1) by $\sigma_x$ to obtain the eigenvalue form

$$\gamma_x \sum_{\mu=t,y} \{\gamma_\mu(i\hbar\partial_\mu + qA_\mu) - \Delta\}\xi_n = i\mu_n \xi_n, \quad (12)$$

where we work in 2+1 dimensions: $\gamma_\mu = (\sigma_z, i\sigma_x, i\sigma_y)$, $A_\mu = (qE_y, 0, 0)$ and the Minkowski metric is used. The operator on the left-hand side is anti-hermitian. The $\xi_n(y)$ are the parabolic cylinder solutions of Eq. (1) written with a discrete index and $\mu_n$ are their corresponding eigenvalues $k_x$ in Eq. (2). Now $\mu_n$ can have positive or negative sign, but when $k_x = 0$, we may transpose the delta term and recall that $\Delta$ is positive. So for the zero modes the symmetry is absent. That $\Delta = 0$ does not invalidate the argument; it just simplifies the picture.

Next, as observed by Ishikawa a small adiabatic change of the vector potential $A$ induces a current in the direction transverse to the external electric field [20]. The partition function $Z$ is the determinant of $\gamma^\mu(i\hbar\partial_\mu + eA_\mu) - \Delta \equiv \gamma \cdot D - \Delta$, so the relevant quantity has the symbolic form $\delta \ln Z = \text{Tr}[(\gamma \cdot D - \Delta)^{-1} \gamma_\mu \delta A^\mu]$. The computation of the induced current density $j_x$ (transverse to $E_0$) is carried in the usual way [10], invoking the relation between the zero and non-zero modes,

$$j_x = e\hbar \left( \gamma_x \frac{1}{\gamma_\mu D_\mu - \Delta} \right) = e \,\text{tr} \left( \frac{1}{i\hbar\partial_x + \gamma_x \sum_{\mu = t, y} \{\gamma_\mu(i\hbar\partial_\mu + eA_\mu) - \Delta\}} \right)$$

$$= e \int_{-\infty}^{\infty} \frac{d\lambda}{2\pi} \int_{-\infty}^{\infty} \frac{d\omega}{2\pi} \sum_n \frac{1}{\lambda + i\mu_n} \text{Tr } e^{i\lambda x/\hbar} \xi_n^+(y - y_0) e^{i\omega t} e^{-i\lambda x/\hbar} \xi_n(y - y_0) e^{-i\omega t} \qquad (13)$$

in which we inserted suitable normalized eigenstates (cf. Eq. (12)) and adapted the discrete index. Replacing $d\omega$ by $\frac{qE_0}{2\pi\hbar} dy_0$ and integrating we find, $j_x = \frac{e^2 E_0}{2\pi^2 \hbar} \int d\lambda \sum \frac{1}{\lambda + i\mu_n}$. If we consider momentarily only the nonzero eigenvalues $\mu_n \neq 0$ the sum can be split into four sums: $\Sigma_{\mu_n>0} \frac{1}{\pm\lambda\pm i\mu_n}$. Thus for $\mu_n \neq 0$ the net result is $j_x = 0$. Extra care is required for the zero modes since there are only two (not four) terms of the form $\frac{1}{\lambda\pm i\Delta}$ (when $\Delta = 0$ two solutions are still present because the negative energy solution has the opposite charge to its counterpart). The net result is thus not zero but, $j_x = i\frac{e^2}{2\pi\hbar}E_0$ or $\frac{e^2}{4\pi\hbar}\varepsilon^{x\nu\rho}F_{\nu\rho}$, when expressed in the usual form of the axial anomaly. The spectral asymmetry is clearly responsible for this result. It is not an accident that the coefficient of $E_0$ is precisely the vacuum Hall conductance $\sigma_{xy}^{vac}$ which can be shown to be due to the polarization of the Dirac sea as a response to the electric field [21]. Note that $j_x$ is even under parity, whereas $E_0$ is odd [22]. Unlike the particle-hole pair undergoing zitterbewegung this polarization phenomenon arising from the zero modes does not lead to a measurable dipole moment in graphene.

In summary, a moderately strong electric field in the plane of graphene is energetically capable of creating one particle-hole pair per cell in graphene. As each charge undergoes zitterbewegung they are separated in the presence of the field; the resulting dipole moment per cell is predicted to be of similar strength as that for typical BN nanotubes, and hence detectable experimentally. The zero modes on the other hand induce an anomalous current transverse to the electric field, but in contrast with the particle-hole pairs, do not lead to separated charge carriers. This phenomenon is peculiar to monolayer graphene and does not have a counterpart in bilayer graphene.

The support of NRF/NUS (Grant No. R-143-000-357-281) and Agency for Science, Technology and Research (ASTAR) is gratefully acknowledged.

1. J. Reinhardt, B. Muller, and W. Greiner, Phys. Rev. A **24**, 103 (1981).

2. B. Muller, H. Peitz, J. Rafelski, and W. Greiner, Phys. Rev. Lett. **28**, 1235 (1972); Y. Hirata and H. Minakata, Phys. Rev. D **34**, 2493 (1986).

3. A. K. Geim and K. S. Novoselov, Nat. Mater. **6**, 183 (2007); A. Geim and A. H. MacDonald, Physics Today **60**, 35 (2007).

4. Pair production has a long history going back to J. Schwinger, Phys. Rev. **82**, 664 (1951).

5. M. V. Krasin'kova and A. P. Paugurt, Tech. Phys. Lett. **31**, 316 (2005).

6. E. J. Mele and P. Kral, Phys. Rev Lett. **88**, 056803 (2002).

7. J. Schliemann, D. Loss, and R. M Westervelt, Phys. Rev. B. **73**, 085323 (2006).

8. T. Rusin and W. Zawadzki, Phys. Rev. B **76**, 195439 (2007).

9. B. M. Garraway and K. A. Suominen, Rep. Prog. Phys. **58**, 365, (1995).

10. K. Ishikawa, Phys. Rev. D **31**, 1432 (1985).

11. A. S. Blaer, N. H. Christ and J.-F. Tang, Phys. Rev. Lett. **47**, 1364 (1981)

12. M. Abramowitz and I. Stegun, *Handbook of Mathematical Functions* (Dover, New York, 1964).

13. S. P. Kim and D. N. Page, Phys. Rev. D **65**, 105002 (2002).

14. A. Hansen and F. Ravndal, Phys. Scr. **23**, 1033 (1981); T. Padmanabhan, Pramana **37**, 179 (1991).

15. S. Coleman, *Aspects of Symmetry* (Cambridge U. Press, Cambridge, 1985).

16. J. Gonzalez, F. Guinea and M. A. H. Vozmediano, Nucl. Phys. **B406**. 771 (1993).

17. E. McCann and V. I. Fal'ko, Phys. Rev. Lett. **96**, 086805 (2006).

18. E. Schrodinger, Sitzungsber. Preuss. Akad. Wiss. Phys. Math. Kl. **24**, 418 (1930).

19. J. A. Lock, Am. J. Phys. **47**, 797 (1979).

20. K. Ishikawa, Phys. Rev. Lett. **53**, 1615 (1984).

21. N. Fumita and K. Shizuya, Phys. Rev. **D 49**, 4277 (1994).

22. A. Abouelsaood, Phys. Rev. Lett. 54, 1973 (1985).

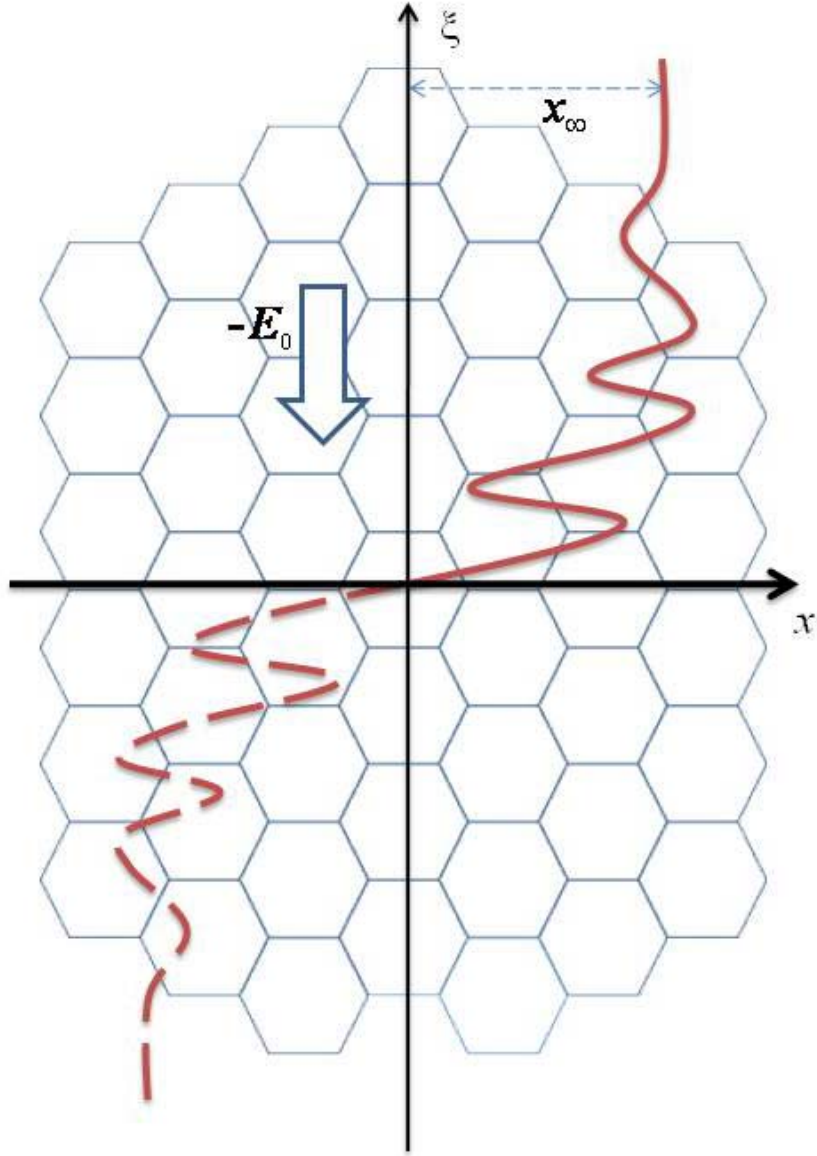

FIG. 1 Schematic diagram of a static electric field $E_0$ (orientation is arbitrary) in the plane of monolayer graphene. A particle-hole pair created at $(x = 0, \xi = 0)$ is separated by the field and as they undergo zitterbewegung (solid path for the particle and dashed path for the hole), they evolve into a separated pair with transverse separation $2x_\infty$. This can be measured as a transverse polarization.

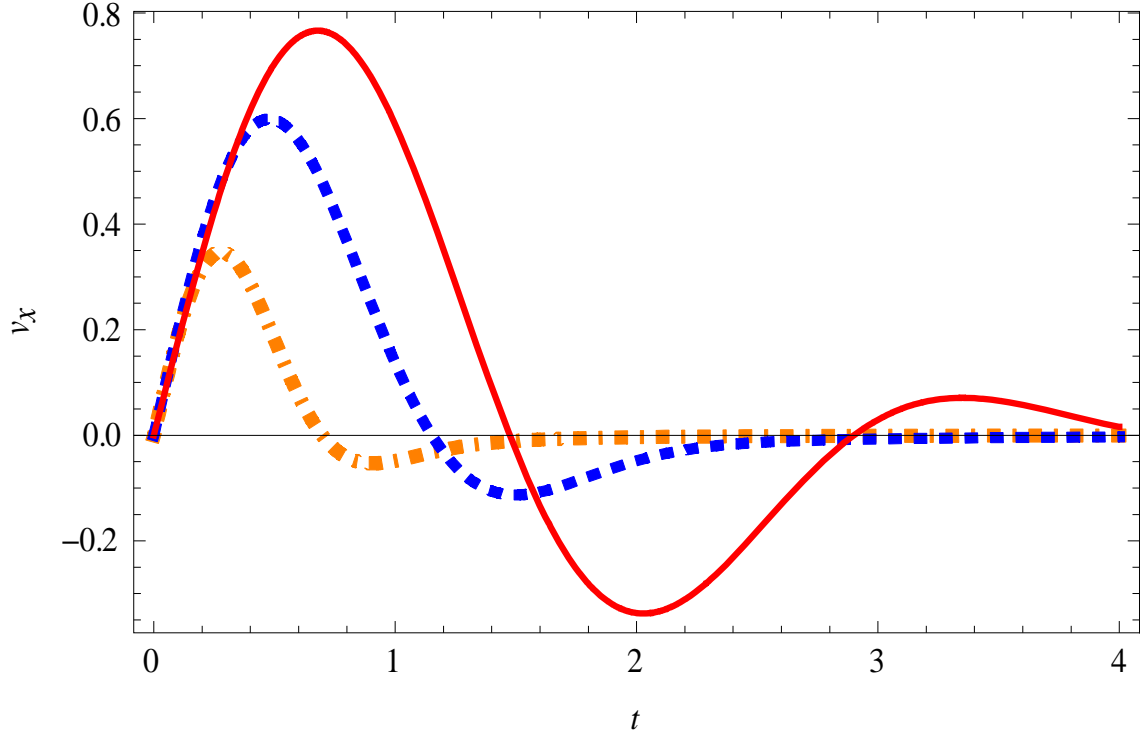

FIG 2. Plots of $<v_x>$ (in units of $v_F = 10^6 m/s$) versus time (in fs) for $k_{0y} = 10^9 m^{-1}$, $k_{0x} = 0$, $d = 5$ (dot-dashed, orange), 10 (dashed, blue), 20 (red) Å. The electric field is $E_0 = 10^7 V\ m^{-1}$. The oscillations are strongly damped, and depend on the width $d$ of the Gaussian wave-packet.